\documentclass{article}
\usepackage{amsfonts}

\newtheorem{theorem}{Theorem}

\newtheorem{corollary}[theorem]{Corollary}

\newtheorem{definition}[theorem]{Definition}
\newtheorem{example}[theorem]{Example}

\newtheorem{proposition}[theorem]{Proposition}
\newtheorem{remark}[theorem]{Remark}

\newenvironment{proof}[1][Proof]{\noindent\textbf{#1.} }{\ \rule{0.5em}{0.5em}}

\newcommand{\cB}{{\mathcal B}}

\newcommand{\sC}{{\mathbb C}}

\newcommand{\sT}{{\mathbb T}}

\newcommand{\sZ}{{\mathbb Z}}

\newcommand{\spec}{\mbox{\rm spec}}

\begin{document}

\title{Essential spectra and exponential estimates of eigenfunctions
 of lattice operators of quantum mechanics}
\author{Vladimir Rabinovich (Instituto Politecnico Nacional, Mexico) \and %
Steffen Roch (Technical University of Darmstadt, Germany)}
\date{}
\maketitle

\begin{abstract}
This paper is devoted to estimates of the exponential decay of
eigenfunctions of difference operators on the lattice ${\mathbb{Z}}^n$ which
are discrete analogs of the Schr\"{o}dinger, Dirac and square-root
Klein-Gordon operators. Our investigation of the essential spectra and the 
exponential decay of eigenfunctions of the discrete spectra is based on the
calculus of the so-called pseudodifference operators (i.e.,
pseudodifferential operators on the group ${\mathbb{Z}}^n$ with analytic
symbols, as developed in the paper \cite{RRActa}), and the limit operators
method (see \cite{RRSB} and the references cited there). We obtain
a description of the location of the essential spectra and estimates of the
eigenfunctions of the discrete spectra of the main lattice operators of
quantum mechanics, namely: matrix Schr\"{o}dinger operators on ${\mathbb{Z}}%
^n$, Dirac operators on ${\mathbb{Z}}^3$, and square root Klein-Gordon
operators on ${\mathbb{Z}}^n$.
\end{abstract}


\section{Introduction}

The main goal of this paper is estimates of the exponential decay of
eigenfunctions of difference operators on the lattice ${\mathbb{Z}}^n$.
Particular attention is paid to the discrete analogs of the Schr\"{o}dinger,
Dirac and square-root Klein-Gordon operators. Schr\"{o}dinger operators on
the lattice ${\mathbb{Z}}^n$ appear at several places, so in the
tight binding model in solid state physics, in the propagation of
spin waves and waves in quasi-crystals, and in mathematical models of
nano-structure, to mention only a few (see, \cite{Mattis,Mogilner,SR3,Teschl}).
There is an extensive bibliography devoted to different aspects of the
spectral theory of discrete Schr\"{o}dinger operators. Let us only mention
the papers \cite{AlbaverioLakaev,AlbLakMum,LakaevMuminov,LastSimon1,Shubindis,Teschl,Yafaev},
and see also the references cited there. In our recent papers \cite{RRJOP1,RRJOP2},
we study the essential spectra of discrete Schr\"{o}dinger operators with variable
magnetic and electric potentials on the lattice ${\mathbb{Z}}^n$ and on
periodic combinatorial graphs by means of the limit operators method (for the latter see also
\cite{RRSB}).

In the last time, also discrete relativistic operators attracted many attention. They were used, e.g., in comparative studies of relativistic
and nonrelativistic electronlocalization phenomena \cite{2}, in relativistic investigations of electrical conduction in disordered systems \cite{21}, in the construction of supertransparent models with supersymmetric structures \cite{25}, and in relativistic tunnelling problems \cite{20}.

The problem of exponential estimates of solutions of the elliptic
partial differential equations with applications to the Schr\"{o}dinger
operator is a classical one. There are an extensive bibliography devoted to this
problem (see for instance \cite{Ag1,Ag,Buzano,FrHerbst,FrHerbstHof,HS}).
Exponential estimates of solutions of pseudodifferential equations are
considered in \cite{LR,Mar,Mar1,Nakam,RTUB,RZ}. Note also the our recent
papers \cite{RRG,RROT} where we proposed a new approach to exponential
estimates for partial differential and pseudodifferential operators, based
on the limit operators method (see \cite{RRSB}).

Our approach to study essential spectra and the exponential decay of
eigenfunctions is based on the calculus of pseudodifference operators (i.e.,
pseudodifferential operators on the group ${\mathbb{Z}}^n$) with analytic symbols as
developed in \cite{RRActa}, and on the limit operators method (see \cite{RRSB}
and the references cited there).

The paper is organized as follows. In Section 2 we recall some auxiliary
facts on the pseudodifference operators with analytic symbols on ${\sZ}^n$,
limit operators, essential spectra and the behavior of solutions of
pseudodifference equations at infinity.

In Section 3 we consider the discrete Schr\"{o}dinger operators on $l^{2}(\sZ^n, \, \sC^N)$ of the form
\[
(Hu)(x) = \sum_{k=1}^n \left(V_{e_{k}} - e^{ia_k(x)}\right) \left(V_{-e_k} -
e^{-ia_k(x)}\right) u(x) + \Phi(x) u(x)
\]
where $V_{e_k}$ is the operator of shift by $e_k$, the $a_k$ are real-valued
bounded slowly oscillating functions on ${\mathbb{Z}}^n$, and $\Phi $ is a
Hermitian slowly oscillating and bounded matrix function on ${\mathbb{Z}}^n$. 
We show that the essential spectrum $\mathrm{sp}_{ess} \, H$ of $H$ is the
interval
\[
\mathrm{sp}_{ess} \, H = \bigcup_{j=1}^n [\lambda_j^{\inf}, \,
\lambda_j^{\sup} + 4n]
\]
where
\[
\lambda_j^{\inf} := \liminf_{x \to \infty} \lambda_j (\Phi(x)), \qquad
\lambda_j^{\sup} := \limsup_{x \to \infty} \lambda_j (\Phi(x))
\]
and where the $\lambda_j (\Phi (x))$ are the increasingly ordered eigenvalues
of the matrix $\Phi (x)$, i.e.
\[
\lambda_1 (\Phi (x)) < \lambda_2 (\Phi (x)) < \ldots < \lambda_N (\Phi (x))
\]
for $x\in {\mathbb{Z}}^n$ large enough. Note that $\mathrm{sp}_{ess} \, H$
does not depend on the exponents $a_k$, and that there is a gap
$(\lambda_j^{\sup} + 4n, \, \lambda_{j+1}^{\inf})$ in the essential spectrum
of $H$ if $\lambda_j^{\sup} + 4n < \lambda_{j+1}^{\inf}$.

We also obtain the following estimates of eigenfunctions belonging to points in the discrete
spectrum of $H$. In each of the cases

\begin{itemize}
\item $\lambda \in (\lambda_j^{\sup} + 4n, \, \lambda_{j+1}^{\inf})$ is an
eigenvalue of $H$ and
\[
0 < r < \cosh^{-1} \left( \frac{\min \{\lambda - \lambda_j^{\sup} - 2n, \,
\lambda_{j+1}^{\inf} - \lambda +2n \}}{2n}\right),
\]

\item $\lambda > \lambda_N^{\sup} + 4n$ is an eigenvalue of $H$ and
\[
0 < r < \cosh^{-1} \left( \frac{\lambda - \lambda_N^{\sup} - 2n}{2n}
\right),
\]

\item $\lambda < \lambda_1^{\inf}$ is an eigenvalue of $H$ and
\[
0 < r < \cosh^{-1} \left( \frac{\lambda_1^{\inf} - \lambda +2n}{2n} \right),
\]
\end{itemize}
every $\lambda$-eigenfunction $u$ of $H$ has the property that $e^{r|x|} u \in l^p({%
\mathbb{Z}}^n, \, {\mathbb{C}}^N)$ for every $1 < p < \infty$.

In Section 4 we introduce self-adjoint Dirac operators on the lattice
${\mathbb{Z}}^3$ with variable slowly oscillating electric potentials. In
accordance with the general properties of Dirac operators on ${\mathbb{R}}^3$
(see for instance \cite{Bjork,Taller}), the corresponding discrete Dirac
operator on ${\mathbb{Z}}^3$ should be a self-adjoint system of the first
order difference operators. We thus let
\begin{equation}  \label{I1}
{\mathcal{D}} := {\mathcal{D}}_0 + e \Phi E_4
\end{equation}
where
\[
{\mathcal{D}}_0 := c \hbar d_k \gamma^k + c^2m \, \gamma^0,
\]
$E_N$ is the $N \times N$ unit matrix, the $\gamma^k$ with $k=0, 1, 2, 3$
refer to the $4 \times 4$ Dirac matrices, the
\[
d_k := I - V_{e_{k}}, \quad k=1,2,3
\]
are difference operators of the first order, $\hbar$ is Planck's constant,
$c$ is the speed of light, $m$ and $e$ are the mass and the charge of the
electron and, finally, $\Phi$ is the real electric potential. The operator
${\mathcal{D}}$, acting on $l^2 ({\mathbb{Z}}^3, \, {\mathbb{C}}^4)$, can be
considered as the direct discrete analog of the Dirac operator on
${\mathbb{R}}^3$, but note that ${\mathcal{D}}$ is not \emph{self-adjoint} on
$l^2 ({\mathbb{Z}}^3, \, {\mathbb{C}}^4)$. To force the self-adjointness, we
consider the "symmetrization" ${\mathbb{D}} := {\mathbb{D}}_0 + e \Phi I$ of
${\mathcal{D}}$ with
\begin{equation}  \label{1}
{\mathbb{D}}_0 := \left(
\begin{array}{cc}
0 & {\mathcal{D}}_0 \\
{\mathcal{D}}_0^* & 0%
\end{array}
\right),
\end{equation}
which acts on $l^2 ({\mathbb{Z}}^3, \, {\mathbb{C}}^8)$. The operator
${\mathbb{D}}$ is self-adjoint, and
\[
{\mathbb{D}}_0^2 = \left(
\begin{array}{cc}
(\hbar^2 c^2 \Gamma + m^2 c^4) E_4 & 0 \\
0 & (\hbar^2 c^2 \Gamma + m^2 c^4) E_4%
\end{array}
\right),
\]
where $\hbar^2 c^2 \Gamma + m^2 c^4$ is the lattice Klein-Gordon Hamiltonian
with Laplacian
\[
\Gamma := \sum_{k=1}^3 d_k^* d_k = \sum_{k=1}^3 (2I - V_{e_k} -
V_{e_k}^*).
\]
Note that one-dimensional Dirac operators of the form (\ref{1}) on
${\mathbb{Z}}$ were considered in \cite{OP1,OP2}.

We prove that the essential spectrum of ${\mathbb{D}}$ is the union
\begin{eqnarray*}
\mathrm{sp}_{ess} \, {\mathbb{D}} &=& [e \Phi^{\inf} - \sqrt{12 \hbar^2 c^2
+ m^2 c^4}, \, e \Phi^{\sup} - mc^2] \\
&& \quad \cup \; [e \Phi^{\inf} + mc^2, \, e \Phi^{\sup} + \sqrt{12 \hbar^2 c^2
+ m^2 c^4}],
\end{eqnarray*}
where
\[
\Phi^{\inf} := \liminf_{x \to \infty} \Phi (x), \qquad \Phi^{\sup} :=
\limsup_{x \to \infty} \Phi (x).
\]
Again we observe that if $e\Phi^{\sup} - e \Phi^{\inf} < 2 m c^2$, then the
essential spectrum of ${\mathbb{D}}$ has the gap $(e \Phi^{\sup} - mc^2, \,
e \Phi^{\inf} + mc^2)$.

We also obtain the following estimates of eigenfunctions of the discrete
spectrum. Let $\lambda $ be a point of the discrete spectrum, and let 
$\lambda$ and $r>0$ satisfy one of the conditions
\begin{itemize}
\item
$\lambda \in (e \Phi^{\sup} - mc^2, \, e \Phi^{\inf} + mc^2)$ and
\[
0 < r < \cosh^{-1} \left( \frac{m^2 c^4 - \max \left\{ (e \Phi^{\inf} -
\lambda)^2, \, (e \Phi^{\sup} - \lambda)^2 \right\} +6 \hbar^2 c^2} {6 \hbar^2 c^2} \right);
\]
\item
$\lambda > e \Phi^{\sup} + \sqrt{12 \hbar^2 c^2 + m^2 c^4}$ and
\[
0 < r < \cosh^{-1} \left( \frac{(e \Phi^{\sup} - \lambda)^2 - m^2 c^4 - 6
\hbar^2 c^2}{6 \hbar^2 c^2}\right);
\]
\item
$\lambda < e \Phi^{\inf} - \sqrt{12 \hbar^2 c^2 + m^2 c^4}$ and
\[
0 < r < \cosh^{-1} \left( \frac{(e \Phi^{\inf} - \lambda)^2 - m^2 c^4 - 6
\hbar^2 c^2}{6 \hbar^2 c^2}\right).
\]
\end{itemize}
Then every $\lambda$-eigenfunction $u$ of the operator ${\mathbb{D}}$
satisfies $e^{r|x|} u \in l^p ({\mathbb{Z}}^3, \, {\mathbb{C}}^8)$ for every $p
\in (1, \infty)$.

In Section 5, we consider the lattice model of the relativistic square root
Klein-Gordon operator as the pseudodifference operator of the form
\[
{\mathcal{K}} := \sqrt{c^2 \hbar^2 \Gamma + m^2 c^4} + e \Phi
\]
on $l^2({\mathbb{Z}}^n)$. We determine the essential spectrum of ${\mathcal{K}}$ 
and obtain exact estimates of the exponential decay at infinity of eigenfunctions 
of the discrete spectrum.

\section{Pseudodifference operators, essential spectra, and exponential
estimates}

\subsection{Some function spaces}

For each Banach space $X$, $\mathcal{B}(X)$ refers to the Banach algebra of all
bounded linear operators acting on $X$. For $1 \le p \le \infty$, we let 
$l^p ({\mathbb{Z}}^n, \, {\mathbb{C}}^N)$ denote the Banach space of all
functions on ${\mathbb{Z}}^n$ with values in ${\mathbb{C}}^N$ with the norm
\[
\|f\|_{l^p ({\mathbb{Z}}^n, \, {\mathbb{C}}^N)}^p := \sum_{x \in {\mathbb{Z}}%
^n} \|f(x)\|_{{\mathbb{C}}^N}^p < \infty \quad \mbox{if} \; p < \infty,
\]
\[
\|f\|_{l^\infty ({\mathbb{Z}}^n, \, {\mathbb{C}}^N)} := \sup_{x \in {\mathbb{%
Z}}^n} \|f(x)\|_{{\mathbb{C}}^N} < \infty.
\]
The choice of the norm on ${\mathbb{C}}^N$ is not of importance in general;
only for $p=2$ we choose the Euclidean norm (such that $l^2 ({\mathbb{Z}}^n,
\, {\mathbb{C}}^N)$ becomes a Hilbert space and ${\mathcal{B}}({\mathbb{C}}^N)$
a $C^*$-algebra in the usual way). Given a positive function $w$ on ${\mathbb{Z}}^n$,
which we will call a weight, let $l^p ({\mathbb{Z}}^n, \, {\mathbb{C}}^N, \, w)$
stand for the Banach space of all functions on ${\mathbb{Z}}^n$ with values
in ${\mathbb{C}}^N$ such that
\[
\|u\|_{l^p ({\mathbb{Z}}^n, \, {\mathbb{C}}^N, \, w)} := \|wu\|_{l^p ({%
\mathbb{Z}}^n, \, {\mathbb{C}}^N)} < \infty.
\]
Similarly, we write $l^\infty ({\mathbb{Z}}^n, {\mathcal{B}} (\mathbb{C}^N))$ for the
Banach algebra of all bounded functions on ${\mathbb{Z}}^n$ with values in ${\mathcal{B}}
({\mathbb{C}}^N)$ and the norm
\[
\|f\|_{l^\infty ({\mathbb{Z}}^n, \, {\mathcal{B}} ({\mathbb{C}}^N))} :=
\sup_{x \in {\mathbb{Z}}^n} \|f(x)\|_{{\mathcal{B}} ({\mathbb{C}}^N)} <
\infty.
\]
Finally, we call a function $a \in l^\infty ({\mathbb{Z}}^n, \, {\mathcal{B}}
({\mathbb{C}}^N))$ \emph{slowly oscillating} if
\[
\lim_{x \to \infty} \|a(x+y) - a(x)\|_{{\mathcal{B}}({\mathbb{C}}^N)} = 0
\]
for every point $y \in {\mathbb{Z}}^n$. We denote the class of all slowly
oscillating functions by $SO({\mathbb{Z}}^n, \, {\mathcal{B}}({\mathbb{C}}%
^N))$ and write simply $SO({\mathbb{Z}}^n)$ in case $N=1$.

\subsection{Pseudodifference operators}

Consider the $n$-dimensional torus ${\mathbb{T}}^n$ as a multiplicative group and
let
\[
d\mu := \left( \frac{1}{2\pi i}\right)^n \frac{dt_1 \cdot \ldots \cdot dt_n}{%
t_1 \cdot \ldots \cdot t_n} = \left( \frac{1}{2\pi i}\right)^n \frac{dt}{t}
\]
denote the corresponding normalized Haar measure on ${\mathbb{T}}^n$.

\begin{definition}
\label{dp3} Let ${\mathcal{S}}(N)$ denote the class of all matrix-valued
functions $a = (a_{ij})_{i,j=1}^n$ on ${\mathbb{Z}}^n \times {\mathbb{T}}^n$
with
\begin{equation}  \label{p5}
\|a\|_k := \sup_{(x,t) \in {\mathbb{Z}}^n \times {\mathbb{T}}^n, \, |\alpha|
\le k} \|\partial_t^\alpha a(x,t)\|_{{\mathcal{B}}({\mathbb{C}}^N)} < \infty
\end{equation}
for every non-negative integer $k$, provided with the convergence defined by
the semi-norms $|a|_k$. To each function $a \in {\mathcal{S}}(N)$, we
associate the \emph{pseudodifference operator}
\begin{equation}  \label{p6}
(\mbox{\rm Op} \, (a) u)(x) := \int_{{\mathbb{T}}^n} a(x,t) \hat{u}(t) t^x
\, d\mu (t), \quad x \in {\mathbb{Z}}^n,
\end{equation}
which is defined on vector-valued functions with finite support. Here,
$\hat{u}$ refers to the discrete Fourier transform of $u$, i.e.,
\[
\hat{u}(t) := \sum_{x \in \mathbb{Z}^n} u(x) t^x, \quad t \in \mathbb{T}^n.
\]
We denote the class of all pseudodifference operators by $OP{\mathcal{S}} (N)$.
\end{definition}

Pseudodifference operators on ${\mathbb{Z}}^n$ can be thought of as the discrete
analog of pseudodifferential operators on ${\mathbb{R}}^n$ (see for instance
\cite{Shubin,Taylor}); they can be also interpreted as (abstract)
pseudodifferential operators with respect to the group ${\mathbb{Z}}^n$. For
another representation of pseudodifference operators, we need the
operator $V_\alpha$ of shift by $\alpha \in {\mathbb{Z}}^n$, i.e. the
operator $V_\alpha$ on $l^p ({\mathbb{Z}}^n, \, {\mathbb{C}}^N)$ which acts via
\[
(V_\alpha u)(x) = u (x - \alpha), \quad x \in {\mathbb{Z}}^n.
\]
Then the operator $\mbox{\rm Op} \, (a)$ can be written as
\begin{equation}  \label{p6'}
\mbox{\rm Op} \, (a) = \sum_{\alpha \in {\mathbb{Z}}^n} a_\alpha V_\alpha
\end{equation}
where
\[
a_\alpha (x) := \int_{{\mathbb{T}}^n} a(x,t) t^\alpha \, d\mu(t).
\]
Integrating by parts we obtain
\begin{equation}  \label{dp7}
\|a_\alpha\|_{l^\infty ({\mathbb{Z}}^n, {\mathcal{B}}({\mathbb{C}}^N))} \le
C |a|_2 (1 + |\alpha|)^{-2},
\end{equation}
whence
\begin{equation}  \label{dp7'}
\|\mbox{\rm Op} \, (a) \|_{W({\mathbb{Z}}^n, {\mathbb{C}}^N)} :=
\sum_{\alpha \in {\mathbb{Z}}^n} \|a_\alpha\|_{l^\infty ({\mathbb{Z}}^n, {%
\mathcal{B}} ({\mathbb{C}}^N))} < \infty.
\end{equation}
We thus obtain that the pseudodifference operator $\mbox{\rm Op} \,
(a)$ belongs to the Wiener algebra $W({\mathbb{Z}}^n, {\mathbb{C}}^N)$
which, by definition, consists of all operators of the form (\ref{p6'}) with
norm (\ref{dp7'}). It is an immediate consequence of this fact that all
operators $\mbox{\rm Op} \, (a)$ in $OP {\mathcal{S}}(N)$ are bounded on
$l^p ({\mathbb{Z}}^n, {\mathbb{C}}^N)$ for all $p \in [1, \infty]$.
Moreover, since the algebra $W({\mathbb{Z}}^n, {\mathbb{C}}^N)$ is inverse
closed in ${\mathcal{B}}(l^p ({\mathbb{Z}}^n, {\mathbb{C}}^N))$, the
spectrum of $\mbox{\rm Op} \, (a) \in OP {\mathcal{S}}(N)$ is independent of
the underlying space $l^p ({\mathbb{Z}}^n, {\mathbb{C}}^N)$.

Note that the operator (\ref{p6}) can be also written as
\[
\mbox{\rm Op} \, (a) u(x) = \sum_{y \in {\mathbb{Z}}^n} \int_{{\mathbb{T}}%
^n} a(x,t) t^{x-y} u(y) \, d\mu (t),
\]
which leads to the following generalization of pseudodifference operators.
Let $a$ be a function on ${\mathbb{Z}}^n \times {\mathbb{Z}}^n \times {%
\mathbb{T}}^n$ with values in ${\mathcal{B}}({\mathbb{C}}^N)$ which is
subject to the estimates
\begin{equation}  \label{dp9}
|a|_k = : \sup_{(x,y,t) \in {\mathbb{Z}}^n \times {\mathbb{Z}}^n \times {%
\mathbb{T}}^n, \, |\alpha| \le k} \|\partial_t^\alpha a(x,y,t)\|_{{\mathcal{B%
}}({\mathbb{C}}^N)} < \infty
\end{equation}
for every non-negative integer $k$. Let ${\mathcal{S}}_d (N)$ denote the set
of all functions with these properties. To each function $a \in {\mathcal{S}}%
_d (N)$, we associate the \emph{pseudodifference operator with double symbol}
\begin{equation}  \label{dp10}
(\mbox{\rm Op}_d \, (a) u)(x) := \sum_{y \in {\mathbb{Z}}^n} \int_{{\mathbb{T%
}}^n} a(x,y,t) u(y) t^{x-y} \, d\mu (t)
\end{equation}%
where $u: {\mathbb{Z}}^n \to {\mathbb{C}}^N$ is a function with finite
support. The right-hand side of (\ref{dp10}) has to be understood as in
(5.6) in \cite{RRSB}, which is in analogy with the definition of an
oscillatory integral (see \cite{Shubin} and also Section 4.1.2 in \cite{RRSB}).
The class of all operators of this form is denoted by $OP {\mathcal{S}}_d
(N)$.

The representation of operators on ${\mathbb{Z}}^n$ as pseudodifference
operators is very convenient due to the fact that one has explicit formulas for
products and adjoints of such operators. The basic results are as follows.

\begin{proposition}
\label{pp2} $(i)$ Let $a, \, b \in {\mathcal{S}} (N)$. Then the product $%
\mbox{\rm Op} \, (a) \mbox{\rm Op} \, (b)$ is an operator in $OP {\mathcal{S}%
}(N)$, and $\mbox{\rm Op} \, (a) \mbox{\rm Op} \, (b) = \mbox{\rm Op} \, (c)$
with
\begin{equation}  \label{p11}
c(x,t) = \sum_{y \in {\mathbb{Z}}^n} \int_{{\mathbb{T}}^n} a(x,t\tau) b(x+y,
\tau) \tau^{-y} \, d\mu (\tau),
\end{equation}
with the right-hand side understood as an oscillatory integral.
\\[1mm]
$(ii)$ Let $a \in {\mathcal{S}} (N)$ and consider $\mbox{\rm Op} \, (a)$ as
acting on $l^p ({\mathbb{Z}}^n, \, {\mathbb{C}}^N)$ with $p \in (1, \infty)$.
Then the adjoint operator of $\mbox{\rm Op} \, (a)$ belongs to $OP {%
\mathcal{S}} (N)$, too, and it is of the form $\mbox{\rm Op} \, (a)^* = %
\mbox{\rm Op} \, (b)$ with
\begin{equation}  \label{p12}
b(x,t) = \sum_{y\in {\mathbb{Z}}^n} \int_{{\mathbb{T}}^n} a^* (x+y, t\tau)
\tau^{-y} \,d\mu (\tau ),
\end{equation}
where $a^*(x,t)$ is the usual adjoint (i.e., transposed and complex conjugated)
matrix. \\[1mm]
$(iii)$ Let $a \in {\mathcal{S}}_d (N)$. Then $\mbox{\rm Op}_d \, (a) \in OP
{\mathcal{S}} (N)$, and $\mbox{\rm Op}_d \, (a) = \mbox{\rm Op} \, (a^{\#})$
where
\[
a^{\#}(x,t) = \sum_{y\in {\mathbb{Z}}^n} \int_{{\mathbb{T}}^n} a(x+y,t\tau)
\tau^{-y} \, d\mu (\tau).
\]
\end{proposition}

\subsection{Limit operators and the essential spectrum}

Recall that an operator $A \in {\mathcal{B}} (X)$ is a \emph{Fredholm
operator} if its kernel $\ker A = \{ x \in X : Ax = 0 \} $ and its cokernel $%
\mbox{\rm coker} \, A = X/(AX)$ are finite-dimensional linear spaces. The essential
spectrum of $A$ consists of all points $\lambda \in {\mathbb{C}}$ such that
the operator $A - \lambda I$ is not a Fredholm operator. We denote the
(usual) spectrum and the essential spectrum of $A$ by $\mbox{\rm spec}_X A$
and $\mathrm{sp}_{ess} \,_X A$, respectively.

Our main tool to study the Fredholm property is limit operators. The
following definition is crucial in what follows.

\begin{definition}
Let $A \in {\mathcal{B}} (l^p ({\mathbb{Z}}^n, \, {\mathbb{C}}^N))$ with $p
\in (1, \infty)$, and let $h : {\mathbb{N}} \to {\mathbb{Z}}^n$ be a
sequence which tends to infinity in the sense that $|h(n)| \to \infty$ as
$n \to \infty$. An operator $A^h \in {\mathcal{B}} (l^p ({\mathbb{Z}}^n, \, {%
\mathbb{C}}^N))$ is called a limit operator of $A$ with respect to the
sequence $h$ if
\[
\mbox{\rm s-lim}_{m \to \infty} V_{-h(m)} A V_{h(m)} = A^{h} \quad \mbox{and}
\quad \mbox{\rm s-lim}_{m \to \infty} V_{-h(m)} A^* V_{h(m)} = (A^h)^*,
\]
where $\mbox{\rm s-lim}$ refers to the strong limit. Clearly, every operator has
at most one limit operator with respect to a given sequence. We denote the
set of all limit operators of $A$ by $op(A)$.
\end{definition}

Let $aI$ be the operator of multiplication by the function $a \in l^\infty ({%
\mathbb{Z}}^n, {\mathcal{B}}({\mathbb{C}}^N))$. A standard Cantor diagonal
argument shows that every sequence $h$ tending to infinity possesses a
subsequence $g$ such that, for every $x \in {\mathbb{Z}}^n$, the limit
\[
\lim_{m \to \infty} a(x + g(m)) =: a^g(x)
\]
exists. Clearly, $a^g$ is again in $l^\infty ({\mathbb{Z}}^n, {\mathcal{B}}
({\mathbb{C}}^N))$. Hence, all limit operators of $aI$ are of the form $a^g I$.
In particular, if $a \in SO({\mathbb{Z}}^n, {\mathcal{B}} ({\mathbb{C}}%
^N)) $, then it follows easily from the definition of a slowly oscillating
function that all limit operators of $aI$ are of the form $a^g I$ where now 
$a^g \in {\mathcal{B}}({\mathbb{C}}^N)$ is a constant function.

Let $\mbox{\rm Op} \, (a) \in OP {\mathcal{S}}(N)$, and let $h : {%
\mathbb{N}} \to {\mathbb{Z}}^n$ be a sequence tending to infinity. Then $%
V_{-h(m)} A V_{h(m)} = \mbox{\rm Op} \, (a_m)$ with $a_m(x) := a(x + h(m),t)$%
. It follows as above that the sequence $h$ has a subsequence $g$ such that $a(x
+ g(m),t)$ converges to a limit $a^g (x,t)$ for every $x \in {\mathbb{Z}}^n$
uniformly with respect to $t \in {\mathbb{T}}^n$. One can prove that the
so-defined function $a^g$ belongs to ${\mathcal{S}}(N)$ and the associated operator
$\mbox{\rm Op} \, (a^g)$ is the limit operator of $\mbox{\rm Op} \, (a)$ with
respect to $g$.

The following theorem gives a complete description of the essential
spectrum of pseudodifference operators in terms of their limit operators.

\begin{theorem}
\label{tp3} Let $a \in {\mathcal{S}}(N)$. Then, for every $p \in (1, \infty)$,
\begin{equation}  \label{p13}
\mathrm{sp}_{ess} \,_{l^p} \mbox{\rm Op} \, (a) = \bigcup_{\mbox{\rm \scriptsize Op} \,
(a^g) \in op (A)} \mbox{\rm spec}_{l^r} \mbox{\rm Op} \, (a^g)
\end{equation}
where $r \in [1, \infty]$ is arbitrary.
\end{theorem}

Since $\mbox{\rm spec}_{l^r} \mbox{\rm Op} \, (a^g)$ does not depend on the
underlying space, the essential spectrum $\mathrm{sp}_{ess} \,_{l^p} %
\mbox{\rm Op} \, (a)$ is independent of $p \in (1, \infty)$. Hence, in what
follows we will omit the explicit notation of the underlying space in the
spectrum and the essential spectrum.

\subsection{Pseudodifference operators with analytic symbols and exponential
estimates of eigenfunctions}

For $r > 1$ let ${\mathbb{K}}_r$ be the annulus $\{t \in {\mathbb{C}} : r^{-1} < |t| < r
\}$, and let ${\mathbb{K}}_r^n$ be the product ${\mathbb{K}}_r \times \ldots
\times {\mathbb{K}}_r$ of $n$ factors.

\begin{definition}
\label{dp4} Let ${\mathcal{S}} (N, \, {\mathbb{K}}_r^n)$ denote the set of
all functions
\[
a : {\mathbb{Z}}^n \times {\mathbb{K}}_r^n \to {\mathcal{B}}({\mathbb{C}}^N)
\]
which are analytic with respect to $t$ in the domain ${\mathbb{K}}_r^n$ and
satisfy the estimates
\[
|a|_k := \sum_{|\alpha| \le k} \sup_{x \in {\mathbb{Z}}^n, \, t \in {\mathbb{%
K}}_r^n} \| \partial_t^\alpha a(x,t)\|_{{\mathcal{B}}({\mathbb{C}}^N)} <
\infty
\]
for every non-negative integer $k$. With every function $a \in {\mathcal{S}}
(N, \, {\mathbb{K}}_r^n)$, we associate a pseudodifference operator defined
on vector-valued functions with finite support via $(\ref{p6})$, and we
denote the corresponding class of pseudodifference operators by $OP {%
\mathcal{S}} (N, \, {\mathbb{K}}_r^n)$.
\end{definition}

\begin{definition}
\label{dp5} For $r > 1$, let ${\mathcal{W}} ({\mathbb{K}}_r^n)$ denote the class of all
exponential weights $w = \exp v$, where $v$ is the restriction onto ${%
\mathbb{Z}}^n$ of a function $\tilde{v} \in C^{(1)} ({\mathbb{R}}^n)$ with
the property that, for every point $x \in {\mathbb{R}}^n$ and every $j=1,...,n$,
\begin{equation}  \label{3.8}
-\log r < \frac{\partial \tilde{v}(x)}{\partial x_{j}} < \log r.
\end{equation}
\end{definition}

In what follows we will denote both the function $\tilde{v}$ on ${\mathbb{R}}%
^n$ and its restriction onto ${\mathbb{Z}}^n$ by $v$. Note that it is an
immediate consequence of Definition \ref{dp5} that if $w \in {\mathcal{W}} ({%
\mathbb{K}}_r^n)$, then $w^\mu \in {\mathcal{W}} ({\mathbb{K}}_r^n)$ for
every $\mu \in [-1, \, 1]$.

\begin{proposition}
\label{pp3} Let $A := \mbox{\rm Op} \, (a) \in OP {\mathcal{S}}(N, {\mathbb{K%
}}_r^n)$ and $w \in {\mathcal{W}} ({\mathbb{K}}_r^n)$. Then the operator $%
A_w := wAw^{-1}$, defined on vector-valued functions with finite support,
belongs to the class $OP {\mathcal{S}}_d (N)$, and $A_w = \mbox{\rm Op}_d \,
(b)$ with
\[
b(x,y,t) = a(x, e^{-\theta_w (x,y)} \cdot t)
\]
where
\[
e^{-\theta_w (x,y)} \cdot t := \left( e^{-\theta_{w,1} (x,y)} t_1, \,
e^{-\theta_{w,2} (x,y)} t_2, \, \ldots, \, e^{-\theta_{w,n}(x,y)} t_n
\right)
\]
and
\[
\theta_{w,j} (x,y) := \int_0^1 \frac{\partial v((1-\gamma )x+\gamma y)}{%
\partial x_j} \, d\gamma.
\]
\end{proposition}

Proposition \ref{pp2} and estimate (\ref{dp8}) imply the following theorem.

\begin{theorem}
\label{tp4} Let $a \in {\mathcal{S}} (N, {\mathbb{K}}_r^n)$ and $w \in {%
\mathcal{W}} ({\mathbb{K}}_r^n)$. Then $\mbox{\rm Op} \, (a)$ is a bounded
operator on each of the spaces $l^p ({\mathbb{Z}}^n, \, {\mathbb{C}}^N, w)$
with $1 \le p\le \infty$.
\end{theorem}

Next we consider essential spectra of pseudodifference operators on weighted
spaces. Let $a, \, A$ and $A_w$ be as in Proposition \ref{pp3}. One can
easily check that for $h \in {\mathbb{Z}}^n$
\[
V_{-h} A_w V_h = \mbox{\rm Op}_d \, (b_h) \quad \mbox{with} \quad b_h(x,y,t)
= a(x+h, e^{-\theta_w (x+h, y+h)} \cdot t)).
\]%
Let now $h : {\mathbb{N}} \to {\mathbb{Z}}^n$ be a sequence tending to
infinity. Then there exists a subsequence $g$ of $h$ such that the limit
operator of $A_w$ with respect to $g$ exists and
\begin{equation}  \label{p12'}
A_w^g = \mbox{\rm Op}_d \, (b^g) \quad \mbox{with} \quad b^g (x,y,t) = a^g
(x, e^{\theta_w^g (x,y)} \cdot t)
\end{equation}
where
\begin{equation}  \label{12''}
a^g (x,t) := \lim_{m \to \infty} a(x + g(m), t)
\end{equation}
and
\begin{equation}  \label{031208.1}
\theta_w^g (x,y) := \lim_{m \to \infty} \int_0^1 \nabla v((1-\gamma )x +
\gamma y + g(m)) \, d\gamma.
\end{equation}
The limits in (\ref{12''}) and (\ref{031208.1}) are understood as pointwise
with respect to $x, y \in {\mathbb{Z}}^n$ and uniform with respect to $t \in
{\mathbb{T}}^n$.

\begin{theorem}
\label{tp5} Let $a \in {\mathcal{S}} (N, {\mathbb{K}}_r^n)$ and $w \in {%
\mathcal{W}} ({\mathbb{K}}_r^n)$, set $A := \mbox{\rm Op} \, (a)$ and $A_w
:= w A w^{-1}$, and consider $A$ as operating from $l^p ({\mathbb{Z}}^n,
{\mathbb{C}}^N, w)$ to $l^p ({\mathbb{Z}}^n, {\mathbb{C}}^N, w)$ where $p \in
(1, \infty)$. Then
\[
\mathrm{sp}_{ess} \, \mbox{\rm Op} \, (a) = \bigcup_{\mbox{\rm \scriptsize Op}_d \,
(b^g) \in op (A_w)} \mbox{\rm spec} \, \mbox{\rm Op}_d \, (b^g)
\]
with $b^g$ as in $(\ref{p12'})$.
\end{theorem}

\begin{remark}
\label{r1} 
Note that the essential spectrum of an operator in $OP {\mathcal{S}}(N, {\mathbb{K}}_r^n)$,
considered as acting on $l^p ({\mathbb{Z}}^n, {\mathbb{C}}^N, w)$, is
independent of $p \in (1, \infty)$, but it can depend on the weight in general. But if the
weight $w = e^v$ has the property that
\begin{equation} \label{r2}
\lim_{x \rightarrow \infty} \nabla v(x) = 0,
\end{equation}
then the symbol $a_w^g$ does not depend on the weight and, hence, the essential
spectrum of $\mbox{\rm Op} \, (a) \in \cB (l^p ({\mathbb{Z}}^n, 
{\mathbb{C}}^N, w))$ with $a \in {\mathcal{S}} (N, {\mathbb{K}}_r^n)$ is 
independent both on $p \in (1, \infty)$ and on the weight $w$. Important 
examples of weights satisfying $(\ref{r2})$ are the power weights
$w(x) = (1 + |x|^2)^{s/2} = e^{\frac{s}{2} log(1 + |x|^2)}$ with $s>0$
and the subexponential weights $w(x) = e^{\alpha|x|^{\beta}}$ where
$\alpha>0$ and $\beta \in (0,1)$.
\end{remark}

The next theorem provides exponential estimates of solutions of
pseudodifference equations.

\begin{theorem}
\label{tp6} Let $A = Op(a) \in OP {\mathcal{S}} (N, {\mathbb{K}}_r^n)$ and $%
w \in {\mathcal{W}} ({\mathbb{K}}_r^n)$. Suppose that $\lim_{x \to \infty}
w(x) = +\infty$ and that $0$ is not in the essential spectrum of $A_{w^\mu}
: l^p({\mathbb{Z}}^n, {\mathbb{C}}^N) \to l^p({\mathbb{Z}}^n, {\mathbb{C}}%
^N) $ for some $p \in (1, \infty)$ and every $\mu \in [-1, 1]$. If $u \in
l^p ({\mathbb{Z}}^n, {\mathbb{C}}^N, w^{-1})$ is a solution of the equation $%
Au=f$ with $f \in l^p ({\mathbb{Z}}^n, {\mathbb{C}}^N, w)$, then $u \in l^p (%
{\mathbb{Z}}^n,{\mathbb{C}}^N,w)$.
\end{theorem}

Theorem \ref{tp6} has some important corollaries.

\begin{theorem}
\label{tp7} Let $a,\,A$ and $w$ be as in the previous theorem, and let $%
\lambda $ be an eigenvalue of $A$ which is not in the essential spectrum of $%
A_{w^{\mu }}:l^{p}({\mathbb{Z}}^{n},{\mathbb{C}}^{N})\rightarrow l^{p}({%
\mathbb{Z}}^{n},{\mathbb{C}}^{N})$ for some $p\in (1,\infty )$ and every $%
\mu \in \lbrack 0,1]$. Then every $\lambda $-eigen\-function belongs to
$l^{p}({\mathbb{Z}}^{n}, {\mathbb{C}}^{N},w)$ for every $p \in (1, \infty)$.
\end{theorem}

\begin{corollary}
Let $A = Op(a) \in OP {\mathcal{S}}(N, {\mathbb{K}}_r^n)$ and let $\lambda$
be an eigenvalue of $A$ which is not in the essential spectrum of $A$. Then
every $\lambda $-eigenfunction $u=(u_{1},...,u_{N})$ satisfies the sub-exponential
estimate
\begin{equation}
\sup |u_{i}(x)| \le C_{i} e^{-\alpha|x|^{\beta}}, \quad x \in \mathbb{Z}^{n}, \, 
i=1,...,N
\label{eqv1}
\end{equation}
for arbitrary $\alpha>0$ and $0<\beta<1$.
\end{corollary}
\begin{proof}
Let $w(x)=e^{v(x)}$ where $v(x)=\alpha|x|^{\beta}$ with $\alpha>0$
and $0<\beta<1$. Then $\lim_{x \rightarrow \infty} \nabla v(x) = 0$, whence
$A^{g}_{w^{\mu}} = A^{g}$ for every limit operator $A^{g}$. Let
$\lambda $ be an eigenvalue of $A$ which is not in the essential
spectrum of $A$. Then $\lambda$ is not in the essential spectrum of 
$A_{w^{\mu}}$ for every $\mu \in [0,1]$. Hence, by Theorem \ref{tp7}, 
every $\lambda$-eigenfunction belongs to each of the spaces
$l^{p}({\mathbb{Z}}^{n},{\mathbb{C}}^{N},w)$ with $p\in (1, \infty)$.
Applying the H\"{o}lder inequality we obtain estimate (\ref{eqv1}).
\end{proof}

We are now going to specialize these results to the context of slowly
oscillating symbols and slowly oscillating weights.

\begin{definition}
\label{dp6} The symbol $a \in {\mathcal{S}}(N, {\mathbb{K}}_r^n)$ is said to
be \emph{slowly oscillating} if
\[
\lim_{x \to \infty} \sup_{t \in {\mathbb{T}}^n} \|a(x+y,t) - a(x,t)\|_{{%
\mathcal{B}}({\mathbb{C}}^N)} = 0
\]
for every $y \in {\mathbb{Z}}^n$. We write ${\mathcal{S}}^{sl} (N, {\mathbb{K%
}}_r^n)$ for the class of all slowly oscillating symbols and $OP
{\mathcal{S}}^{sl} (N, {\mathbb{K}}_r^n)$ for the corresponding class of
pseudodifference operators.
\end{definition}

\begin{definition}
\label{dp8} The weight $w = e^v \in {\mathcal{W}} ({\mathbb{K}}_r)$ is \emph{%
slowly oscillating} if the partial derivatives $\frac{\partial v}{\partial x_j}$
are slowly oscillating for $j = 1, \, \ldots, \, n$. We denote the class of all
slowly oscillating weights by ${\mathcal{W}}^{sl} ({\mathbb{K}}_r)$.
\end{definition}

\begin{example}
If $v(x) = \gamma |x|$, then $\frac{\partial v(x)}{\partial x_{j}} = \gamma
\frac{x_{j}}{|x|}$ for $j=1, \, \ldots, \, n$. Thus, $w := e^v$ is in ${\mathcal{W}}
^{sl} ({\mathbb{K}}_r)$ if $\gamma < r$.
\end{example}

The next theorem describes the structure of the limit operators of the operator 
$A_w = wAw^{-1}$ if $A \in OP{\mathcal{S}}^{sl} (N, {\mathbb{K}}_r^n)$ and $w
\in {\mathcal{W}}^{sl} ({\mathbb{K}}_r)$.

\begin{theorem}
\label{tp9} Let $A = \mbox{\rm Op} \, (a) \in OP {\mathcal{S}}^{sl} (N, {%
\mathbb{K}}_r^n)$ and $w \in {\mathcal{W}}^{sl} ({\mathbb{K}}_r)$. Then the
limit operator $A_w^g$ of $A_w$ with respect to the sequence $g$ tending to infinity
exists if the limits 
\begin{equation}  \label{p15}
a_g(t) = \lim_{m \to \infty} a (g(m), t), \qquad \theta_w^g = \lim_{m \to
\infty} (\nabla v)(g(m))
\end{equation}
exist. In this case, it is of the form 
\begin{equation}  \label{p14}
A_w^g = \mbox{\rm Op} \, (c_g) \quad \mbox{with} \quad c_g (x,t) = a_g
(\theta_w^g \cdot t).
\end{equation}
\end{theorem}

Consequently, if $A$ and $w$ are as in this theorem, then the limit
operators $A_{w^\mu}^g$ are invariant with respect to shifts. This fact
implies the following explicit description of their essential spectra.
Let $\{ \lambda_j (A_{w^\mu}^g)(t) \}_{j=1}^n$ denote the eigenvalues of the
matrix $a_g (\theta_{w^\mu}^g \cdot t)$. Then
\[
\mbox{\rm spec}_{l^p ({\mathbb{Z}}^n, {\mathbb{C}}^N)} A_{w^\mu}^g =
\bigcup_{j=1}^N \{ \lambda_j(A_{w^\mu}^g)(t) : t \in \sT^n \; \mbox{and} \; j = 1, \ldots, n \},
\]
whence
\[
\mathrm{sp}_{ess} \,_{l^p ({\mathbb{Z}}^n, {\mathbb{C}}^N)} A_{w^\mu} =
\bigcup_{A_{w^\mu}^g \in op (A_{w^\mu})} \bigcup_{j=1}^N \{ \lambda_j (A_{w^\mu}^g)(t) : t \in
\sT^n \; \mbox{and} \; j = 1, \ldots, n \}.
\]

\section{Matrix Schr\"{o}dinger operators}

\subsection{Essential spectrum}

%
In this section we consider the essential spectrum and the behavior at infinity
of eigenfunctions of general discrete Schr\"{o}dinger operators acting on
$u \in l^2 ({\mathbb{Z}}^n, {\mathbb{C}}^N)$ by
\begin{equation}  \label{s1}
(Hu)(x) = \sum_{k=1}^n (V_{e_k} - e^{ia_k(x)}) \, (V_{-e_k} - e^{-ia_k(x)})
u(x) + \Phi(x)
\end{equation}
where $e_k = (0, \ldots, 0, 1, 0, \ldots,0)$ with the 1 standing at the $k$th 
place, the $a_k \in SO({\mathbb{Z}}^n)$ are real-valued, and $\Phi \in
SO({\mathbb{Z}}^n, {\mathcal{B}}({\mathbb{C}}^N))$ is
Hermitian. The vector $a := (a_1, \ldots, a_n)$ is the discrete
analog of the magnetic potential, whereas $\Phi $ can be viewed of as a
discrete analog of the electric potential. Since the essential
spectrum of $H$ is independent of $p \in (1, \infty)$, we consider
the case $p=2$ only. Note that our assumptions guarantee that $H$ is a
self-adjoint operator on $l^2 ({\mathbb{Z}}^n, {\mathbb{C}}^N)$.

The limit operators $H_g$ of $H$ are of form
\begin{eqnarray*}
H_g & = & \sum_{k=1}^n (V_{e_k} - e^{ia_k^g} I) \, (V_{-e_k} -
e^{-ia_k^g}I) + \Phi^g I \\
& = & \sum_{k=1}^n (2 I - e^{-ia_k^g} V_{-e_k} - e^{ia_k^g} V_{e_k} ) + \Phi^g I
\end{eqnarray*}
with the constant functions
\[
a_k^g = \lim_{m \to \infty} a_k (x + g(m)) \quad \mbox{and} \quad \Phi^g = 
\lim_{m \to \infty} \Phi (x + g(m)).
\]
Let $U:l^2 ({\mathbb{Z}}^n, {\mathbb{C}}^N) \to l^2 ({\mathbb{Z}}^n, {%
\mathbb{C}}^N)$ be the unitary operator
\[
(Uu)(x) = e^{-i \langle a^{g}, x \rangle} u(x), \quad a^g = (a_1^g, \ldots,
a_n^g).
\]
Then
\[
U^* H_g U = \sum_{k=1}^n (2 I - V_{-e_k} - V_{e_k}) + \Phi^g.
\]
Further, the operator $H_g^\prime := U^* H_g U$ is unitarily equivalent to the
operator of multiplication by the function
\[
\tilde{H}_g(\psi_1, \, \ldots, \, \psi_n) := 4 \sum_{k=1}^n \sin^2 \frac{%
\psi_k}{2} + \Phi^g, \quad \psi_k \in [0, \, 2 \pi],
\]
acting on $L^2 ([0, \, 2\pi]^n, \, {\mathbb{C}}^N)$. Hence,
\[
\mbox{\rm spec} \, H_g = \mbox{\rm spec} \, H_g^\prime = \bigcup_{j=1}^N
[\lambda_j (\Phi^g), \, \lambda_j (\Phi^g) + 4n],
\]
where the $\lambda_j (\Phi^g)$ refer to the eigenvalues of the matrix $%
\Phi^g $. Applying formula (\ref{p13}) we obtain
\begin{equation}  \label{s2}
\mathrm{sp}_{ess} \, H = \bigcup_g \bigcup_{j=1}^N [\lambda_j (\Phi^g), \,
\lambda_j (\Phi^g) + 4n]
\end{equation}
where the first union is taken over all sequences $g$ for which the limit
operator of $H$ exists. Let $\lambda_j (\Phi(x))$, $j=1, \ldots, N$, denote
the eigenvalues of the matrix $\Phi(x)$. We suppose that these eigenvalues
are simple for $x$ large enough and that they are increasingly ordered,
\[
\lambda_1 (\Phi(x)) < \lambda_2 (\Phi(x)) < \ldots < \lambda_N (\Phi(x)).
\]
Then one can show that the functions $x \mapsto \lambda_j (\Phi(x))$ belong
to $SO({\mathbb{Z}}^n)$. Let
\[
\lambda_j^{\inf} := \liminf_{x \to \infty} \lambda_j (\Phi(x)), \qquad
\lambda_j^{\sup} := \limsup_{x \to \infty} \lambda_j (\Phi(x)).
\]
Since the set of the partial limits of a slowly oscillating function on 
${\mathbb{Z}}^n$ is connected for $n > 1$ (see \cite{RRSB}, Theorem 2.4.7), we
conclude from (\ref{s2}) that
\begin{equation}  \label{s2'}
\mathrm{sp}_{ess} \, H = \bigcup_{j=1}^N [\lambda_j^{\inf}, \,
\lambda_j^{\sup} + 4n]
\end{equation}
for $n > 1$. Note that if $\lambda_j^{\sup} + 4n < \lambda_{j+1}^{\inf}$,
then there is the gap $(\lambda_j^{\sup} + 4n, \, \lambda_{j+1}^{\inf})$
in the essential spectrum of $H$.

In case $n=1$, the set of the partial limits of a slowly oscillating function
on ${\mathbb{Z}}$ consists of two connected components, which collect the
partial limits as $x \to - \infty$ and $x \to + \infty$, respectively. Accordingly, 
in this case we set
\[
\lambda_j^{\inf, \pm} := \liminf_{x \to \pm \infty} \lambda_j (\Phi(x)), \qquad
\lambda_j^{\sup, \pm} := \limsup_{x \to \pm \infty} \lambda_j (\Phi(x)).
\]
and obtain
\[
\mathrm{sp}_{ess} \, H = \bigcup_{j=1}^N \left( [\lambda_j^{\inf,-}, \,
\lambda_j^{\sup,-} + 4] \cup [\lambda_j^{\inf,+}, \,
\lambda_j^{\sup,+} + 4] \right).
\]

\subsection{Exponential estimates of eigenfunctions}

Our next goal is to apply Theorem \ref{tp6} to eigenfunctions of (discrete)
eigenvalues of the operator $H$ with slowly oscillating potentials. We will
formulate the results for $n > 1$ only; for $n=1$ the non-connectedness of 
the set of the partial limits requires some evident modifications. 
According to (\ref{s2'}), the discrete spectrum of $H$ is located outside
the set $\mathrm{sp}_{ess} \, H = \bigcup_{j=1}^N [\lambda_j^{\inf}, \,
\lambda_j^{\sup} + 4n]$ if $n > 1$.

Let $\cosh^{-1} : [1, +\infty) \to [0, +\infty )$ refer to the function
inverse to $\cosh : [0, +\infty )\to [1, +\infty )$, i.e.,
\[
\cosh^{-1} \mu = \log (\mu + \sqrt{\mu^2 - 1}).
\]
Further let ${\mathcal{R}}^{sl} := \bigcup_{r > 1} {\mathcal{W}} ({\mathbb{%
K}}_r^n)$.

\begin{theorem}
\label{ts1} Let $w = e^v$ be a weight in ${\mathcal{R}}^{sl}$ with $\lim_{x
\to \infty} v(x) = \infty$. Further let $\lambda $ be an eigenvalue of $H$
such that $\lambda \notin \mathrm{sp}_{ess} \, H$ and assume that one of the
following conditions is satisfied: \\[1mm]
$(i)$ there is a $j \in \{1, \, \ldots, \, N\}$ such that $\lambda \in
(\lambda_j^{\sup} + 4n, \, \lambda_{j+1}^{\inf})$ and
\begin{equation}  \label{s3}
\limsup_{x \to \infty} \left| \frac{\partial v(x)}{\partial x_k} \right| <
\cosh^{-1} \left( \frac{\min \{\lambda - \lambda_j^{\sup} - 2n, \,
\lambda_{j+1}^{\inf} - \lambda +2n\}}{2n}\right)
\end{equation}
for every $k=1, \, \ldots, \,n$; \\[1mm]
$(ii)$ $\lambda > \lambda_N^{\sup} + 4n$ and
\[
\limsup_{x \to \infty} \left| \frac{\partial v(x)}{\partial x_k} \right| <
\cosh^{-1} \left( \frac{\lambda -\lambda_N^{\sup} - 2n}{2n} \right)
\]
for every $k=1, \, \ldots, \,n$; \\[1mm]
$(iii)$ $\lambda < \lambda_1^{\inf}$ and
\[
\limsup_{x \to \infty} \left| \frac{\partial v(x)}{\partial x_k} \right| <
\cosh^{-1} \left( \frac{\lambda_1^{\inf} - \lambda + 2n}{2n} \right)
\]
for every $k=1, \, \ldots, \,n$. \\[1mm]
Then every $\lambda$-eigenfunction of $H$ belongs to each of the spaces $l^p ({%
\mathbb{Z}}^n, {\mathbb{C}}^N, w)$ with $p \in (1, \infty)$.
\end{theorem}

\begin{proof}
For $\mu \in [0, \, 1]$, let $H_{w^\mu}^\prime := w^\mu H^\prime w^{-\mu}$.
The limit operators $H_{w^\mu}^{\prime g} - \lambda E$ are unitarily
equivalent to the operator of multiplication by the matrix-function
\[
{\mathcal{H}}_{w^\mu}^g (\psi) = \left( -2 \sum_{j=1}^n \cos (\psi_j + i \mu
\theta_j^g) + 2n - \lambda \right) E + \Phi^g
\]
where
\[
\psi = (\psi_1, \, \ldots, \psi_n) \in [0, \, 2 \pi]^n \quad \mbox{and}
\quad \theta_j^g := \lim_{m \to \infty} \frac{\partial v(g(m))}{\partial x_j}%
.
\]
Note that
\[
\mathfrak{R} ({\mathcal{H}}_{w^\mu}^g (\psi)) = \left( -2 \sum_{j=1}^n \cos
\psi_j \cosh \mu \theta_{w,j}^g + 2n-\lambda \right) E + \Phi^g,
\]
where $\theta_{w,j}^g := \left( \frac{\partial v}{\partial x_j} \right)^g$.
It is easy to check that condition (\ref{s3}) implies that $\lambda \notin %
\mbox{\rm spec} \, H_{w^\mu}^g$ for every limit operator $H_{w^\mu}^g$ of $%
H_{w^\mu}$ and every $\mu \in [0, \,1]$. Hence, by Theorem \ref{tp7}, every $%
\lambda$-eigenfunction belongs to $l^p ({\mathbb{Z}}^n, {\mathbb{C}}^N, w)$
for every $1 < p < \infty$.
\end{proof}

\begin{corollary}
\label{cs1} In each of the following cases \\[1mm]
$(i)$ $\lambda \in (\lambda_j^{\sup} + 4n, \, \lambda_{j+1}^{\inf})$ for
some $j \in \{1, \, \ldots, \, N\}$ and
\[
0 < r < \cosh^{-1} \left( \frac{\min \{\lambda -\lambda_j^{\sup} - 2n, \,
\lambda_{j+1}^{\inf} - \lambda + 2n\}}{2n} \right),
\]
$(ii)$ $\lambda > \lambda_N^{\sup} + 4n$ and $0 < r < \cosh^{-1} \left(
\frac{\lambda -\lambda_N^{\sup} - 2n}{2n} \right)$, \\[1mm]
$(iii)$ $\lambda < \lambda_1^{\inf}$ and $0 < r < \cosh^{-1} \left( \frac{%
\lambda_1^{\inf} - \lambda +2n}{2n} \right)$, \\[1mm]
every $\lambda$-eigenfunction of $H$ belongs to $l^p ({\mathbb{Z}}^n, {%
\mathbb{C}}^N, e^{r |x|)})$ for each $p \in (1,\infty)$.
\end{corollary}

\begin{remark}
In the case of the scalar Schr\"{o}dinger operator $(\ref{s1})$ with $\Phi \in
SO({\mathbb{Z}}^n)$, we have
\[
\mathrm{sp}_{ess} \, H = [\Phi^{\inf}, \, \Phi^{\sup} + 4n]
\]
with $\Phi^{\inf} = \liminf_{x \to \infty} \Phi (x)$ and $\Phi^{\sup} =
\limsup_{x \to \infty} \Phi (x)$. If one of the following conditions holds
for an eigenvalue $\lambda$ of $H$: \\[1mm]
$(i)$ $\lambda > \Phi^{\sup} + 4n$ and $0 < r < \cosh^{-1} \left( \frac{%
\lambda -\Phi^{\sup} - 2n}{2n} \right)$, or \\[1mm]
$(ii)$ $\lambda < \Phi^{\inf}$ and $0 < r < \cosh^{-1} \left( \frac{%
\Phi^{\inf} + 2n -\lambda}{2n} \right)$, \\[1mm]
then every $\lambda$-eigenfunction of $H$ belongs to $l^p ({\mathbb{Z}}^n,
{\mathbb{C}}^N, e^{r|x|})$ for each $p \in (1, \infty)$.
\end{remark}

\section{The discrete Dirac operator}

\subsection{The essential spectrum}

On $l^2 ({\mathbb{Z}}^3, \, {\mathbb{C}}^4)$, we consider the Dirac
operators
\begin{equation}  \label{d1}
{\mathcal{D}} := {\mathcal{D}}_0 + e \Phi I \quad \mbox{and} \quad {%
\mathcal{D}}_0 := c \hbar d_k \gamma^k + c^2 m \gamma^0
\end{equation}
where the $\gamma^k$, $k= 0, 1, 2, 3$, refer to the $4 \times 4$ Dirac
matrices, i.e., they satisfy
\begin{equation}  \label{d3}
\gamma^j \gamma^k + \gamma^k \gamma^j = 2\delta_{jk} E_{4}
\end{equation}%
for all choices of $j, k = 0, 1, 2 ,3$ where $E_4$ stands for the $4
\times 4$ identity matrix. Further,
\[
d_k := I - V_{e_k}, \quad k = 1,2,3
\]
are difference operators of the first order, $\hbar$ is Planck's constant,
$c$ the light speed, $m$ and $e$ are the mass and the charge of the
electron, and $\Phi $ is the electric potential. We suppose that the
function $\Phi$ is real-valued and belongs to the space $SO({\mathbb{Z}}^3)$.

It turns out that the operator ${\mathcal{D}}$ is \emph{not self-adjoint}
on $l^2 ({\mathbb{Z}}^3, \, {\mathbb{C}}^4)$. Therefore we introduce
self-adjoint Dirac operators as the matrix operators
\[
{\mathbb{D}} := {\mathbb{D}}_0 + e \Phi I
\quad \mbox{with} \quad {\mathbb{D}}_0 := \left(
\begin{array}{cc}
0 & {\mathcal{D}}_0 \\
{\mathcal{D}}_0^* & 0%
\end{array}
\right),
\]
acting on the space $l^2 ({\mathbb{Z}}^3, \, {\mathbb{C}}^8)$ (i.e. $I$ 
refers now to the identity operator on that space). First we are going to 
determine the spectrum of ${\mathbb{D}}_{0}$. It is
\begin{equation}
(\mathbb{D}_{0}-\lambda I)(\mathbb{D}_{0} + \lambda I)= 
\left(
\begin{array}{cc}
\mathcal{L} (\lambda) & 0 \\
0 & \mathcal{L} (\lambda)
\end{array}
\right)
\label{df4}
\end{equation}
where $\mathcal{L}(\lambda) = \hbar^{2} c^{2} \Gamma + (m^{2} c^{4} -\lambda^{2})I $, and
\[
\Gamma := \sum_{k=1}^{3}d_{k}^{\ast}d_{k} = \sum_{k=1}^{3}(2I-V_{e_{k}}-V_{e_{k}}^{\ast})
\]%
is the discrete Laplacian with symbol
\[
\hat{\Gamma}(\varphi )=\hat{\Gamma}(\varphi _{1},\,\varphi
_{2},\,\varphi _{3})=\sum_{k=1}^{3}(2-2\cos \varphi _{k}),\quad
\varphi _{k}\in \lbrack 0,\,2\pi ].
\]
Similarly, we denote by $\hat{\mathbb{D}}_{0} (\varphi)$ and 
$\hat{\mathcal{L}}(\lambda ,\varphi )$ the symbols of the operators 
$\mathbb{D}_{0}$ and $\mathcal{L}(\lambda )$, respectively. Then
\begin{equation}
(\hat{\mathbb{D}}_{0} (\varphi )-\lambda
E_{8})(\hat{\mathbb{D}}_{0} (\varphi )+\lambda
E_{8})=\hat{\mathcal{L}}(\lambda ,\varphi )E_{8} \label{df5}
\end{equation}%
with the scalar-valued function
\[
\hat{\mathcal{L}}(\lambda ,\varphi )=\hbar
^{2}c^{2}\sum_{k=1}^{3}(2-2\cos \varphi _{k})+m^{2}c^{4}-\lambda
^{2}.
\]
We claim that $\lambda \in \spec \, \mathbb{D}_{0}$ if and only if
there exists a $\varphi_{0} \in [0, 2\pi]^{3}$ such that $\hat{\mathcal{L}}%
(\lambda ,\varphi _{0})=0$. Indeed, let $\lambda \in \spec \, \mathbb{D}_{0}$.
Then there exists a $\varphi_{0} \in [0, 2\pi]^{3}$ such that $\det 
(\hat{\mathbb{D}}_{0}(\varphi _{0})-\lambda E_{8})=0.$ Hence by (\ref{df5}) 
$\hat{\mathcal{L}}(\lambda , \varphi_{0}) = 0$. Conversely, if 
$\hat{\mathcal{L}}(\lambda ,\varphi _{0}) = 0$, then it follows from (\ref{df5}) that 
\[
(\hat{\mathbb{D}}_{0}(\varphi_{0})-\lambda E_{8}) (\hat{\mathbb{D}}_{0}(\varphi_{0}) + 
\lambda E_{8}) = 0.
\]
Hence, $\det (\hat{\mathbb{D}}_{0}(\varphi _{0})-\lambda E_{8}) = 0$, whence
$\lambda \in \spec \, \mathbb{D}_{0}$.

Since the equation $\hat{\mathcal{L}}(\lambda, \varphi) = 0$ has two 
branches of solutions (spectral curves), namely
\[
\lambda _{\pm }(\varphi )=\pm \sqrt{\hbar
^{2}c^{2}\hat{\Gamma}(\varphi )+m^{2}c^{4}}, \quad \varphi \in [0, 2\pi]^{3},
\]
the spectrum of ${\mathbb{D}}_{0}$ is the union
\[
\mbox{\rm spec} \, {\mathbb{D}}_{0}=[-\sqrt{12\hbar ^{2}c^{2}+m^{2}c^{4}}
, \, -mc^{2}]\cup \lbrack mc^{2},\,\sqrt{12\hbar
^{2}c^{2}+m^{2}c^{4}}].
\]
Our next goal is to determine the essential spectrum of ${\mathbb{D}} = 
{\mathbb{D}}_0 + e \Phi I$. All limit operators of ${\mathbb{D}}$ are of
the form ${\mathbb{D}}^g = {\mathbb{D}}_0 + e \Phi^g I$ where $\Phi^g =
\lim_{j \to \infty} \Phi (g(j))$ is the partial limit of $\Phi$
corresponding to the sequence $g : {\mathbb{N}} \to {\mathbb{Z}}^3$ tending
to infinity. By what we have just seen, this gives
\begin{eqnarray*}
\mbox{\rm spec} \, {\mathbb{D}}^g & = & [e \Phi^g - \sqrt{12 \hbar^2 c^2 +
m^2 c^4}, \, e \Phi^g - m c^2] \\
&& \qquad \cup \, [e \Phi^g + mc^2, \, e\Phi^g + \sqrt{12 \hbar^2
c^2 + m^2 c^4}].
\end{eqnarray*}
Since $\mathrm{sp}_{ess} \, {\mathbb{D}} = \cup_g \mbox{\rm spec} \,
{\mathbb{D}}^g$  we obtain
\begin{eqnarray*}
\mathrm{sp}_{ess} \, {\mathbb{D}} & =& [e \Phi^{\inf} - \sqrt{12 \hbar^2 c^2
+ m^2 c^4}, \, e \Phi^{\sup} - mc^2] \\
&& \qquad \cup \, [e \Phi^{\inf} + mc^2, \, e \Phi^{\sup} + \sqrt{12 \hbar^2
c^2 + m^2 c^4}],
\end{eqnarray*}
where
\[
\Phi^{\inf} := \liminf_{x \to \infty} \Phi (x) \quad \mbox{and} \quad
\Phi^{\sup} := \limsup_{x \to \infty} \Phi (x).
\]
In particular, if $e (\Phi^{\sup} - \Phi^{\inf}) < 2 mc^2$, then the interval 
$(e \Phi^{\sup} - m c^{2}, \, e \Phi^{\inf} + mc^2)$ is a gap in the essential
spectrum of ${\mathbb{D}}$.

\subsection{Exponential estimates of eigenfunctions}

The following is the analog of Theorem \ref{ts1}.

\begin{theorem}
\label{ted1} Let $\lambda \notin \mathrm{sp}_{ess} \, {\mathbb{D}}$ be an
eigenvalue of ${\mathbb{D}} : l^p ({\mathbb{Z}}^3, \, {\mathbb{C}}^8) \to
l^p({\mathbb{Z}}^3, \, {\mathbb{C}}^8)$ with $p \in (1, \infty)$. Assume
further that the weight $w = e^v$ is in ${\mathcal{R}}^{sl}$ and that $%
\lim_{x \to \infty} v(x) = \infty$. If one of the conditions \\[1mm]
$(i)$ $\lambda \in (e \Phi^{\sup} - mc^2, \, e \Phi^{\inf} + mc^2)$ and, for
every $j=1, 2, 3$,
\begin{eqnarray}  \label{d5}
\lefteqn{\limsup_{x \to \infty} \left| \frac{\partial v(x)}{\partial x_{j}}
\right|}  \nonumber \\
&& < \; \cosh^{-1} \left( \frac{m^2 c^4 - \max \left\{ (e \Phi^{\inf} -
\lambda)^2, \, (e \Phi^{\sup} - \lambda)^2 \right\} +6 \hbar^2 c^2}{6
\hbar^2 c^2} \right),
\end{eqnarray}
$(ii)$ $\lambda > e \Phi^{\sup} + \sqrt{12 \hbar^2 c^2 + m^2 c^4}$
and, for every $j=1, 2, 3$,
\begin{equation}  \label{d5'}
\limsup_{x \to \infty} \left| \frac{\partial v(x)}{\partial x_j} \right| <
\cosh^{-1} \left( \frac{(e \Phi^{\sup} - \lambda)^2 - m^2 c^4 - 6 \hbar^2 c^2%
}{6 \hbar^2 c^2} \right),
\end{equation}
$(iii)$ $\lambda < e\Phi^{\inf} - \sqrt{12 \hbar^2 c^2 + m^2 c^4}$
and, for every $j=1, 2, 3$,
\begin{equation}  \label{d5''}
\limsup_{x \to \infty} \left| \frac{\partial v(x)}{\partial x_j} \right| <
\cosh^{-1} \left( \frac{(e \Phi^{\inf} - \lambda)^2 - m^2 c^4 - 6 \hbar^2 c^2%
}{6 \hbar^2 c^2} \right),
\end{equation}
is satisfied, then every $\lambda$-eigenfunction of the operator ${\mathbb{D}%
}$ belongs to the space $l^p ({\mathbb{Z}}^3, \, {\mathbb{C}}^8, w)$ for each 
$p \in (1, \, \infty)$.
\end{theorem}

\begin{proof}
We will prove the assertion in case condition $(i)$ is satisfied. The other
cases follow similarly. Further, since the essential spectrum of ${\mathbb{D}%
}$ and the spectra of the associated limit operators do not depend on $p$,
we can assume that $p=2$ in this proof.

Let condition (\ref{d5}) hold, and let $\lambda$ be an eigenvalue in
the gap $(e \Phi^{\sup} - mc^2, \, e \Phi^{\inf} + mc^2)$ of the
essential spectrum. In order to apply Theorem \ref{tp6} to determine
the decaying behavior of the associated eigenfunction $u_\lambda$,
we need estimates of
the spectrum of the limit operators $({\mathbb{D}}_{w^\mu})^g$ of ${\mathbb{D%
}}_{w^\mu} := w^\mu {\mathbb{D}} w^{-\mu}$ for $\mu \in [0, \, 1]$. The
limit operator $(w^\mu V_{e_k} w^{-\mu})^g$ of $w^\mu V_{e_k} w^{-\mu}$ is
of the form
\[
( w^\mu V_{e_k} w^{-\mu})^g = e^{- \mu\left( \frac{\partial v}{\partial x_k}%
\right)^g} V_{e_k}.
\]
Hence,
\begin{equation}  \label{d6}
({\mathcal{D}}_{w^\mu})^g = \sum_{k=1}^3 c \gamma^k (I - e^{-\mu\left( \frac{\partial v}{%
\partial x_k}\right)^g} V_{e_k}) + m c^2 \gamma^0 + e \Phi^g E_4
\end{equation}
where $\left( \frac{\partial v}{\partial x_k}\right)^g = \lim_{m \to
\infty} \frac{\partial v(g(m))}{\partial x_k}$.

Let $\mathbb{D}^{\prime}=\mathbb{D}_{0} - e\Phi I$. The identity
(\ref{d6}) implies that $({\mathbb{D}}_{w^\mu}^{\prime g} - \lambda
I)({\mathbb{D}}_{w^\mu}^{\prime g} + \lambda I)$ is the diagonal matrix
$\mbox{\rm diag} \, (F, \, F)$ with
\[
F := \hbar^2 c^2 \Gamma_{w^\mu}^g + (m^2 c^4 - (e \Phi^g - \lambda)^2) I
\]
and
\[
\Gamma_{w^\mu}^g = \sum_{k=1}^3 \left( 2I - e^{-\left( \frac{\partial v}{\partial
x_k}\right)^g} V_{e_k} - e^{\left( \frac{\partial v}{\partial x_k}
\right)^g} V_{e_k}^* \right).
\]
The operator $\Gamma_{w^\mu}^g$ is unitarily equivalent to the operator of
multiplication by the function
\[
\hat{\Gamma}_{w^\mu}^g (\varphi) = \hat{\Gamma}_{w^\mu}^g (\varphi_1, \,
\varphi_2, \, \varphi_3) = \sum_{k=1}^3 \left( 2 - 2 \cos \left( \varphi_k + i \left(
\frac{\partial v}{\partial x_k} \right)^g \right) \right)
\]
acting on the space $L^2 ([0, \, 2 \pi]^{3})$. Note that
\[
\mathfrak{R} (\hat{\Gamma}_{w^\mu}^g(\varphi)) = 6 - 2 \sum_{j=1}^3 \cos
\varphi_k \cosh \left( \frac{\partial v}{\partial x_k} \right)^g.
\]
Hence, and by condition (\ref{d5}),
\begin{equation}
\label{de8} \mathfrak{R} (\hbar^2 c^2 \hat{\Gamma}_{w^\mu}^g
(\varphi) + m^2 c^4 - (e \Phi^g - \lambda)^2) \neq 0
\end{equation}
for every sequence $g$ defining a limit operator and for every $\mu
\in [0, \, 1]$. The property (\ref{de8}) implies that $\lambda
\notin \mbox{\rm spec} \, {\mathbb{D}}_{w^\mu}^g$ for every limit
operator ${\mathbb{D}}_{w^\mu}^g$ and every $\mu \in [0, \, 1]$. By
Theorem \ref{tp6}, every $\lambda$-eigenfunction belongs to 
$l^p ({\mathbb{Z}}^{3}, \, {\mathbb{C}}^8, \, w)$ for every $p \in (1, \infty)$.
\end{proof}

For the important case of the symmetric weight $w(x) = e^{r |x|}$, we obtain
the following corollary of Theorem \ref{ted1}.

\begin{corollary}
\label{ted1coro} Let $\lambda $ be an eigenvalue of ${\mathbb{D}} : l^p ({%
\mathbb{Z}}^3, \, {\mathbb{C}}^8) \to l^p ({\mathbb{Z}}^3, \, {\mathbb{C}}%
^8) $. If one of the conditions \\[1mm]
$(i)$ $\lambda \in (e \Phi^{\sup} - m c^2, \, e \Phi^{\inf} + mc^2)$ and
\[
0 < r < \cosh^{-1} \left( \frac{m^2 c^4 + 6 \hbar^2 c^2 - \max \left\{ (e
\Phi^{\inf} - \lambda)^2, \, (e \Phi^{\sup} - \lambda)^2 \right\}}{6 \hbar^2
c^2} \right),
\]
$(ii)$ $\lambda > e \Phi^{\sup} + \sqrt{12 \hbar^2 c^2 + m^2 c^4}$ and
\[
0 < r < \cosh^{-1} \left( \frac{(e \Phi^{\sup} - \lambda)^2 - m^2 c^4 - 6
\hbar^2 c^2}{6 \hbar^2 c^2} \right),
\]
$(iii)$ $\lambda < e \Phi^{\inf} - \sqrt{12 \hbar^2 c^2 + m^2 c^4}$ and
\[
0 < r < \cosh^{-1} \left( \frac{(e \Phi^{\inf} - \lambda)^2 - m^2 c^4 - 6
\hbar^2 c^2}{6 \hbar^2 c^2} \right),
\]
is satisfied, then every $\lambda$-eigenfunction of the operator ${\mathbb{D}%
}$ belongs to the space $l^p ({\mathbb{Z}}^3, \, {\mathbb{C}}^8, \,
e^{r|x|}) $ for every $p \in (1, \, \infty)$.
\end{corollary}

\section{The square-root Klein-Gordon operator}

\subsection{The essential spectrum}

Here we consider the square-root Klein-Gordon operator on $l^2({\mathbb{Z}}%
^n)$, that is the operator
\[
K = \sqrt{c^2 \hbar^2 \Gamma + m^2 c^4} + e \Phi
\]
where $m > 0$ is the mass of the particle, $\hbar > 0$ is Planck's constant, $c
> 0$ the light speed, $\Phi \in SO({\mathbb{Z}}^n)$ a scalar potential, and
\[
\Gamma = \sum_{j=1}^n (2I - V_{e_j} - V_{e_j}^*)
\]
is the discrete Laplacian on ${\mathbb{Z}}^n$. The operator $K_0 := \sqrt{%
c^2 \hbar^2 \Gamma + m^2 c^4}$ is understood as the pseudodifference
operator with symbol
\[
k(\tau) = \sqrt{c^2 \hbar^2 \hat{\Gamma}(\tau) + m^2 c^4} \in {\mathcal{S}},
\]
where $\hat{\Gamma}(\tau) = \sum_{j=1}^n (2 - \tau_j - \tau_j^{-1})$ at $%
\tau = (\tau_1, \, \ldots, \, \tau_n)$. Let
\[
\tilde{\Gamma}(\varphi) := \hat{\Gamma} (e^{i \varphi}) = \sum_{j=1}^n (2 -
2 \cos \varphi_j), \quad \varphi = (\varphi_1, \, \ldots, \, \varphi_n) \in
[0, \, 2 \pi]^n.
\]
Every limit operator of $K$ is unitarily equivalent to an operator of
multiplication by a function of the form
\[
\tilde{K}^g (\varphi) = \sqrt{c^2 \hbar^2 \tilde{\Gamma}(\varphi) + m^2 c^4}
+ e \Phi^g \quad \mbox{with} \quad \Phi^g \in {\mathbb{R}}
\]
acting on $L^2 ([0, \, 2 \pi]^n)$. Thus,
\[
\mbox{\rm spec} \, K^g = \bigcup_g \, [m c^2 + e \Phi^g, \, \sqrt{4n c^2
\hbar^2 + m^2 c^4} + e \Phi^g],
\]
where the union is taken with respect to all sequences $g$ tending to
infinity such that the partial limit $\Phi^g := \lim_{m \to \infty} \Phi
(g(m))$ exists. Consequently,
\[
\mathrm{sp}_{ess} \, K = [m c^2 + e \Phi^{\inf}, \, \sqrt{4n c^2 \hbar^2 +
m^2 c^4} + e \Phi^{\sup}]).
\]

\subsection{Exponential estimates of eigenfunctions}

\begin{theorem}
\label{tek2} Let $\lambda $ be an eigenvalue of the square-root Klein-Gordon
operator $K$ such that $\lambda \notin \mathrm{sp}_{ess} \, K$, and let $w =
e^v$ be a weight in ${\mathcal{R}}^{sl}$ with $\lim_{x \to \infty} v(x) =
\infty$. If one of the conditions \\[1mm]
$(i)$ $\lambda > e \Phi^{\sup} + \sqrt{4n \hbar^2 c^2 + m^2 c^4}$ and
\begin{equation}  \label{k'}
\limsup_{x \to \infty} \left| \frac{\partial v(x)}{\partial x_j} \right| <
\cosh^{-1} \left( \frac{m^2 c^4 - (e \Phi^{\sup} - \lambda)^2 + 2n \hbar^2
c^2}{2n \hbar^2 c^2} \right),
\end{equation}
$(ii)$ $\lambda < e \Phi^{\inf} - \sqrt{4n \hbar^2 c^2 + m^2 c^4}$ and
\begin{equation}  \label{k''}
\limsup_{x \to \infty} \left| \frac{\partial v(x)}{\partial x_j} \right| <
\cosh^{-1} \left( \frac{m^2 c^4 - (e \Phi^{\inf} - \lambda)^2 + 2n \hbar^2
c^2}{2n \hbar^2 c^2} \right),
\end{equation}
is satisfied, then every $\lambda$-eigenfunction of $K$ belongs to $l^p ({%
\mathbb{Z}}^n, \, w)$ for every $p \in (1, \, \infty)$.
\end{theorem}

\begin{proof}
The proof proceeds similarly to the proof of Theorem \ref{ted1}. It is based
on the following construction. Let $w = e^v \in {\mathcal{R}}^{sl}$. Then
the limit operator $K_{w^\mu}^g$ is unitarily equivalent to the operator of
multiplication by the function
\[
\tilde{K}_{w^\mu}^g (\varphi) = \sqrt{c^2 \hbar^2 \tilde{\Gamma} (\varphi +
i (\nabla v))^g + m^2 c^4} + e \Phi^g
\]
acting on $L^2 ([0, \, 2 \pi]^n$. Hence,
\begin{eqnarray*}
{\mathcal{L}}_{w^\mu}^g (\varphi, \, \lambda) & := & (\tilde{K}_{w^\mu}^g
(\varphi) - \lambda) \left( \sqrt{c^2 \hbar^2 \tilde{\Gamma} (\varphi +i
(\nabla v))^g + m^2 c^4} - (e \Phi^g - \lambda) \right) \\
& = & c^2 \hbar^2 \tilde{\Gamma} (\varphi + i (\nabla v))^g + m^2 c^4 - (e
\Phi^g - \lambda)^2,
\end{eqnarray*}
and
\[
\mathfrak{R} ({\mathcal{L}}_{w^\mu}^g (\varphi, \, \lambda)) = c^2 \hbar^2
\sum_{j=1}^n \left( 2 - \cos \varphi_j \cosh (\frac{\partial v}{\partial x_j})^g \right)
+ m^2 c^4 - (e \Phi^g - \lambda)^2.
\]
Note that $\mathfrak{R} ({\mathcal{L}}_{w^\mu}^g (\varphi, \, \lambda)) \neq
0$ for every $\lambda$ satisfying condition $(i)$ or $(ii)$. Hence, $\lambda
\notin \mathrm{sp}_{ess} \, K_{w^\mu}$ for every $\mu \in [0, \, 1]$. Thus,
by Theorem \ref{tp6}, every $\lambda$-eigenfunction belongs to the space $%
l^p ({\mathbb{Z}}^n, \, w)$ for all $p\in (1,\infty).$
\end{proof}

Specifying the weight in the previous theorem as $w(x) = e^{r|x|}$, we
obtain the following.

\begin{theorem}
Let $\lambda$ be an eigenvalue of $K$ such that $\lambda \notin \mathrm{sp}%
_{ess} \, K$. If one of the conditions \\[1mm]
$(i)$ $\lambda > e \Phi^{\sup} + \sqrt{4n \hbar^2 c^2 + m^2 c^4}]$ and
\[
0 < r < \cosh^{-1} \left( \frac{m^2 c^4 - (e \Phi^{\sup} - \lambda)^2 + 2n
\hbar^2 c^2}{2n \hbar^2 c^2} \right),
\]
$(ii)$ $\lambda < e \Phi^{\inf} - \sqrt{4n \hbar^2 c^2 + m^2 c^4}$ and
\[
0 < r < \cosh^{-1} \left( \frac{m^2 c^4 - (e \Phi^{\inf} - \lambda)^2 + 2n
\hbar^2 c^2}{2n \hbar^2 c^2} \right)
\]
is satisfied, then every $\lambda$-eigenfunction of $K$ belongs to the space
$l^p ({\mathbb{Z}}^n, \, e^{r|x|})$ for every $p \in (1, \, \infty)$.
\end{theorem}

\end{document}